\documentclass[journal]{IEEEtran}

\input packages-headers

\RequirePackage[top=2.0cm,bottom=2.0cm,left=1.5cm,right=1.5cm,nohead]{geometry}

\definecolor{MyDarkGreen}{rgb}{0.0,0.4,0.0} 
\definecolor{MyDarkRed}{rgb}{0.4,0.0,0.0} 
\renewcommand{\matlabscript}[3]{
    \begin{itemize}
    \item[]\lstinputlisting[language=Matlab,caption=#3,label=#2]{#1}
    \end{itemize}
}

\newcommand{\MATLAB}{MATLAB$^\text{\textregistered}$}

\begin{document}

\renewcommand{\refname}{References}

\let\redHL=\ignore
\let\jnote=\ignore
\let\tnote=\ignore
\let\tsumm=\ignore

\pagestyle{plain}

\title{{\Huge Oscillator-based Ising Machine}}

\author
{
Tianshi Wang and Jaijeet Roychowdhury\\
{\normalsize
Department of Electrical Engineering and Computer Sciences, University of California, Berkeley, CA, USA}\\
Email: \texttt{\{tianshi, jr\}@berkeley.edu}
}

\maketitle
\thispagestyle{plain}

\begin{abstract}
Many combinatorial optimization problems can be mapped to finding the ground
states of the corresponding Ising Hamiltonians.
The physical systems that can solve optimization problems in this way, namely
Ising machines, have been attracting more and more attention recently.
Our work shows that Ising machines can be realized using almost any nonlinear
self-sustaining oscillators with logic values encoded in their phases.
Many types of such oscillators are readily available for large-scale
integration, with potentials in high-speed and low-power operation.
In this paper, we describe the operation and mechanism of oscillator-based
Ising machines.
The feasibility of our scheme is demonstrated through several examples in
simulation and hardware, among which a simulation study reports average
solutions exceeding those from state-of-art Ising machines on a benchmark
combinatorial optimization problem of size 2000.
\end{abstract}

\section{\normalfont {\large Introduction}}\seclabel{intro}

The Ising model, named after physicist Ernest Ising, started as a model for
explaining domain formation in ferromagnets \cite{ising1925beitrag}
Nowadays, it is considered as a promising non-von Neumann architecture for
solving many combinatorial optimization problems
\cite{lucas2013ising,gu2012encoding}.
In these problems, the objective to be minimized is usually formulated
as the energy of a collection of spins $\{s_i\}$, $i=1,\cdots,n$, represented
by a Hamiltonian function:
\begin{equation}\eqnlabel{IsingH0}
	H = \sum_{i} h_{i} s_i + \sum_{i, j} J_{ij} s_i s_j,
\end{equation}
where $s_i \in \{-1,~+1\}$ are binary integers; coefficients $\{J_{ij}\}$ and
$\{h_i\}$ are real numbers.

An equivalent formula of the Hamiltonian can be written as follows.
\begin{equation}\eqnlabel{IsingH}
	H = \sum_{i, j} J_{ij} s_i s_j,
\end{equation}
where $\{s_i\}$'s size is increased by one, with the last one $s_{n+1} \equiv
+1$;
$\{J_{ij}\}$'s dimension is also increased by one, with $J_{n+1,k} = J_{k,n+1}
= h_k/2$ for $k=1,\cdots,n$.

An Ising machine is a physical realization of the Ising model, \ie, it is a
physical system that can minimize the Hamiltonian function defined in
\eqnref{IsingH0} or \eqnref{IsingH}.
Such a system normally has a graph structure, where the vertices/nodes
represent the spins $\{s_i\}$ and the edges encode the coupling coefficients
$\{J_{ij}\}$ and $\{h_i\}$.
The Ising Hamiltonian can normally be mapped to the energy of this physical
system.
Through annealing, once the system's energy is minimized, the nodes encode the
globally optimal spin configuration, \aka, the ground state.

Several schemes have been proposed towards the realization of Ising machines.
Perhaps the best-known example comes from D-Wave Systems
\cite{johnson2011quantum,bian2014Ising}.
Their quantum annealers use superconducting loops as nodes and interconnect
them with Josephson junctions \cite{harris2010flux}.
Their computing environment requires a temperature below 80mK
\cite{johnson2011quantum}.
There is a lot of controversy \cite{ronnow2014defining} around their advantages
over simulated annealing run on classical computers.
It is believed that through a mechanism known as quantum tunnelling they offer
the largest speed-up on problems with rugged energy landscapes
\cite{denchev2016tunneling}.

Classical annealers that do not rely on quantum mechanics to function have also
been reported with good performances.
One type of annealers use time-multiplexed optical parametric oscillators
(OPOs) as the Ising spins, couple them through delay lines, and control the
coupling with an FPGA \cite{marandi2014network,mcmahon2016ScienceIsing100}.
An Ising machine with a size of 2000 has been reported with a high success
probability for solving the MAX-CUT problems
\cite{inagaki2016ScienceIsing2000}.
Similar to OPOs, mechanical parametric oscillators built with MEMS technology
have also been proposed to use in Ising machines
\cite{mahboob2016electromechanical}; no physical realization has been reported
yet.
Nanomagnets with low energy barriers are another candidate for Ising spins
\cite{sutton2017intrinsic}.
They are given a name ``p-bits'', and are shown through computational
studies to be able to minimize energy functions, which stem from not only
combinatorial optimization problems, but also invertible logic computation
\cite{camsari2017pbits}.
Researchers have also been exploring the possibility of implementing Ising
machines with SRAMs in CMOS technology \cite{yamaoka2016IsingCMOS}.
But ``the efficacy in achieving a global energy minimum is limited''
\cite{yamaoka2016IsingCMOS} due to variation.
The speed-up and accuracy reported by \cite{yamaoka2016IsingCMOS} are instead
based on deterministic on-chip computation paired with external random number
generators --- a digital hardware implementation of the simulated annealing
algorithm.
As such, the CMOS-based scheme is not directly comparable to the other Ising
machines described above.

\ignore{
While the theory has been available for decades, the major obstacle of the
practical large-scale implementation of Ising machine is finding the
appropriate underlying logic components.
It also needs to have intrinsic noise, so that it behaves like a binary random
number.
Moreover, the probability distribution of this random number can be controlled
by an external signal, which is from the connection with other units.
A suitable physical substrate with these features appear to be not common.
(A MRAM naturally has bistability, suitable for binary logic. But to use
them in Ising machines, the energy barrier between the bistable states has to
be lowered than the conditions for which the devices are normally designed.
Parametric resonator is used in MEMS and lasers to achieve bistability.)
}

In this paper, we report a new finding: almost any nonlinear oscillator is
suitable for implementing Ising machines.
Nowadays, a broad choice of such oscillators are available from not just CMOS
technology, but also optics, MEMS, spin torque devices, biochemical reaction
networks, \etc, among which many are integrable and highly energy-efficient.
Therefore, such a finding greatly expands the scope of the physical realization
of Ising machines.

The mechanism is based on a common phenomenon observed in almost all nonlinear
oscillators --- injection locking. 
A variant of it --- sub-harmonic injection locking (SHIL) can excite
multiple stable phase-locked responses in oscillators
\cite{NeRoDATE2012SHIL,WaRoDAC2015MAPPforPHLOGON,Wa2017arXivMetronomes}.
For example, when an oscillator is perturbed by an external periodic input at
close to twice its natural frequency, the oscillator's injection-locked
response is bistable.
The bistable states differ only in the phase/timing and it can be proven that
their phase difference is $180^\circ$
\cite{NeRoDATE2012SHIL,WaRoDAC2015MAPPforPHLOGON}.
In this way, almost any oscillator can be used as a binary logic latch, with
the logic value encoded in the phase of oscillation.
Recently, it has been demonstrated that it is feasible to use oscillators and
phase-based encoding to build finite state machines for general-purpose Boolean
computation based on the conventional von Neumann architecture 
\cite{WaRoUCNC2014PHLOGON,RoPHLOGONprocIEEE2015,WaRoDAC2015MAPPforPHLOGON}.

In this paper, we take the oscillator-based Boolean computation idea one step
further.
An oscillator under SHIL can store a binary logic value securely if the
external periodic perturbation is strong.
As we reduce the strength of this perturbation, SHIL becomes weaker.
And because of the intrinsic phase noise, the oscillator will ``degrade'' from
a binary latch to a binary random number generator.
Naturally, we can speculate that when a few such random number generators are
coupled together, their values will prefer certain configurations over others.
By properly designing the coupling between them, we can encode the Ising
Hamiltonian in a network of such coupled oscillators so that the ground state
is the most preferable state with the lowest ``energy''.
In \secref{mechanism}, we expand on this idea and explore the relationship
between the ``energy'' of coupled oscillators and the Ising Hamiltonian.


It is worth noting that computation with coupled oscillators is not a new
topic.
Associated memory arrays made with oscillators have been attracting research
interest for many years
\cite{Hoppensteadt2000synchronization,maffezzoni2015oscArray,Porod2015physical}.
They are shown to be suitable for some specialized non-Boolean computational
tasks, such as image recognition, edge detection, \etc{}
But an Ising machine differs from them in its capabilities in solving general
combinatorial optimization problems and invertible logic problems
\cite{camsari2017pbits}, which are in the Boolean computation domain.
The enabling technique is the use of SHIL to digitize an oscillator's phase,
which has recently been studied in the context of oscillator-based finite state
machines.
Adapting the technique in realizing Ising machines has become a feasible option
and a natural choice.
It is also an important piece in the framework of oscillator-based Boolean
computation.
Indeed, with oscillators, both Ising machines, which are stochatic in nature,
and deterministic digital computation can be implemented with the same type of
devices, possibly on the same chip, opening up many new possibilities in the
design of computer architectures and algorithms.

It is also known that oscillations of neurons in cortical networks play an
important role in many of the functionalities of neural circuits, ranging from
sensory input processing to working memory retention and decision making.
It has long been suspected that the coupled oscillating neurons function as an
optimizer \cite{hoppensteadt2012weakly}.
While associative memory arrays offer a perspective for understanding this
hypophysis, oscillator-based Ising machines offer another one, showing that it
is possible for coupled oscillators to solve for almost arbitrarily complicated
problems.
In fact, Ising machines can encode any Boolean logic function and are Turing
complete \cite{gu2012encoding}.
The bistability or multistability, instead of coming from SHIL, can also come
from delayed coupling \cite{kim1997multistability}.


In the remainder of this paper, we first describe the idea and mechanism of
oscillator-based Ising machines in \secref{mechanism}.
In particular, we show how the global Lyapunov function of the phase macromodel
of coupled oscillators maps to the Ising Hamiltonian.
We also study how variations in the oscillators' central frequencies affect the
performance of the system.
Then in \secref{speed}, we study how the speed of convergence scales with the
size of the optimization problem, which is often considered as a major
attractiveness of Ising machines.
Several examples of the use of oscillator-based Ising machines are shown in
\secref{examples}, including in both combinatorial optimization problems and
invertible logic applications.

\section{Oscillator-based Ising Machines}\seclabel{mechanism}

In this section, we first sketch out the idea of using oscillators to implement
Ising machines, focusing on the intuition behind it.
Then we describe the mechanism more rigorously by deriving the phase macromodel
of coupled oscillators and studying its relationship with the Ising
Hamiltonian.

As mentioned in \secref{intro}, when an oscillator with natural frequency $f_0$
is perturbed by a small periodic external input at $f_1 \approx f_0$, through
injection locking, its response can lock on to the input in both frequency and
phase.
Many natural phenomena result from this mechanism, \eg, metronomes on a same
platform end up ticking in unison (\figref{osc}), fireflies synchronize their
flashes, neurons fire in unison, \etc{}
Sub-harmonic injection locking (SHIL) is a special type of injection locking.
Under SHIL, an oscillator is perturbed by a periodic input at about twice its
natural frequency, \ie, $2f_1$; we call this perturbation a synchronization
signal (SYNC).
The oscillator will then lock to the sub-harmonic of SYNC while developing
bistable phase locks separated by 180$^\circ$.
In this way, it becomes a logic latch that can store a phase-based binary bit.
Together with phase-based combinational logic gates \cite{WaRoUCNC2014PHLOGON},
finite state machines can then be implemented for general-purpose Boolean
computation.
This mechanism is generic in almost all oscillators; the scheme is not specific
to electrical ones.
A broad choice of nonlinear oscillators --- from MEMS oscillators to optical
lasers, from spin-torque nano-oscillators (STNOs) to oscillating biochemical
reactions and neurons --- become potential candidates for Boolean computation
systems (\figref{osc}).

\begin{figure}[htbp!]
    \centering
    {
        \hspace{0em}\epsfig{file=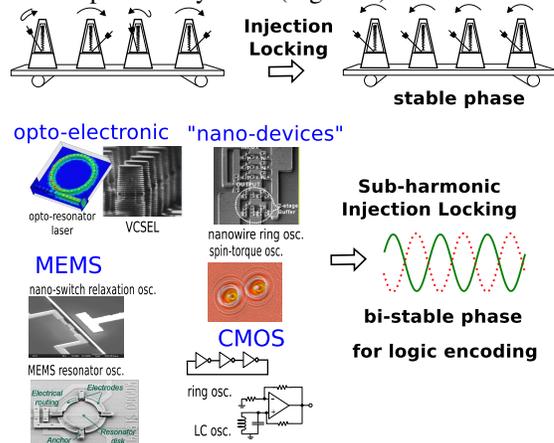, width=0.8\linewidth}
    }
    \caption{ Injection Locking leads to phase lock. In the case of SHIL,
    bistable phase locks enable phase-based logic encoding and storage
    with many types of nano oscillators.
    \figlabel{osc}}
\end{figure}

Furthermore, if we lower SYNC and reduce the effect of SHIL, due to intrinsic
phase noise and detuning, the oscillator will have phase response that
stochastically flips between the bistable states --- it becomes a binary random
number generator.
If two of these oscillators are coupled, \eg, through resistive connection,
instead of having four states \{00\}, \{01\}, \{10\}, \{11\} with even
probabilities, they will have higher probability to end up in \{00\} and \{11\}
--- these two states become the ground states of this simple size-2 Ising
model.
More complicated Ising Hamiltonians can be encoded by coupling more
oscillators, in a similar way to the existing Ising machine proposals.
Specifically, a larger coefficient $J_{ij}$ in \eqnref{IsingH} is represented
with a larger conductance in the resistive connection.

To study this mechanism more properly, we start from the phase macromodel of a
single nonlinear self-sustaining oscillator.
Such an oscillator under perturbation can be described mathematically as a set
of Differential Algebraic Equations (DAEs):
\begin{equation}\eqnlabel{DAE}
	\frac{d}{dt} \vec q(\vec x) + \vec f(\vec x)  + \vec b(t) = \vec 0
\end{equation}
where $\vec x \in \mathbf{R}^n$ are the unknowns in the system, 
$\vec b(t)$ is a small time-varying input.
The oscillator's response can be approximated well as
\begin{equation}\eqnlabel{xt}
	\vec x(t) = \vec x_s(t + \alpha(t))
\end{equation}
where $x_s(t)$ is the oscillator's steady state response without perturbation
($\vec b(t) \equiv \vec 0$);
$\alpha(t)$ is the phase shift caused by the external input and is governed by
the following differential equation:
\begin{equation}\eqnlabel{PPV}
	\frac{d}{dt} \alpha(t) = \vec v^T(t + \alpha(t)) \cdot \vec b(t)
\end{equation}
where the vector $\vec v(t)$ is known as the Perturbation Projection Vector
(PPV) \cite{DeMeRoTCAS2000} of the oscillator.
Assume the oscillator's natural frequency is
$\omega_0 = 2\pi f_0 = 2\pi/T_0$.
Then $\vec v(t)$ is a $T_0$-periodic vector that can be extracted
numerically from the DAEs of the oscillator without knowing any information
about the input $\vec b(t)$.
Put in other words, it is a property intrinsic to the oscillator that 
captures its phase response to small external inputs.
PPV can be used to model and predict injection locking effectively
\cite{LaRoMTT2004}.

When the external perturbation $\vec u(t)$ is periodic itself with frequency
$\omega_1 = 2\pi f_1 = 2\pi /T_1$, equation \eqnref{PPV} can be rewritten as
\begin{equation}\eqnlabel{PPVphi}
\frac{d}{dt} \Delta \phi(t) = \omega_0-\omega_1 + \omega_0 \cdot \vec v_{(2\pi)}^T(\omega_1\cdot t +
\Delta \phi(t)) \cdot \vec u_{(2\pi)}(\omega_1\cdot t),
\end{equation}
where $\Delta \phi(t) = (\omega_0-\omega_1)\cdot t + \omega_0\cdot \alpha(t)$
--- when injection locking occurs, it is the phase difference between the
oscillator's response and the periodic perturbation;
both $\vec v_{(2\pi)}$ and $\vec u_{(2\pi)}$ are $2\pi$-periodic functions ---
$\vec v_{(2\pi)}(t) = \vec v(t/\omega_0)$,
$\vec u_{(2\pi)}(t) = \vec u(t/\omega_1)$.

In a special case, when both $\vec v(t)$ and $\vec u(t)$ are sinusoidal scaler
functions, \ie, we assume $\vec v_{(2\pi)}(t) = A_v\sin(t)$
and $\vec u_{(2\pi)}(t) = A_u\cos(t + \Delta \phi_u)$,
\eqnref{PPVphi} can be rewritten as 
{\small
\begin{align}
\frac{d}{dt} \Delta \phi(t)
=& \omega_0-\omega_1+\omega_0 \cdot
A_v\sin\left(\omega_1 t + \Delta \phi(t)\right)
\cdot
A_u\cos\left(\omega_1 t + \Delta \phi_u\right)\\
=& \omega_0-\omega_1+\omega_0 \cdot [
\frac{1}{2}A_vA_u\sin\left(\Delta \phi(t) - \Delta \phi_u\right) \nonumber\\
&+
\frac{1}{2}A_vA_u\sin\left(2\omega_1 t + \Delta \phi(t) + \Delta \phi_u\right)
].
\end{align}
}

The last term is fast varying with time. If we average it out and let
$A_1=\frac{1}{2}A_vA_u$, we get
\begin{equation}\eqnlabel{Adler}
    \frac{d}{dt} \Delta \phi(t) =
    \omega_0-\omega_1+\omega_0 A_1 \sin\left(\Delta \phi(t) - \Delta \phi_u\right).
\end{equation}

\eqnref{Adler} is known as the Adler's equation \cite{Adler1973}.
It can explain many interesting properties of injection locking.
For example, if we assume there to be no detuning, \ie, $\omega_0=\omega_1$,
the steady state equation $\frac{d}{dt} \Delta \phi(t) = 0$ has two sets of
solutions --- $\phi_u + 2k\pi$ and $\phi_u + 2k\pi + \pi$,
where $k\in \mathrm{Z}$.
We can linearize the system around the solutions and analyze the slopes, which
indicate the stability of these solutions.
Results show that the latter sets are stable.
This indicates that the oscillator's phase will be stably locked to the input
with only one possible phase shift value under injection locking, which matches
observation.
Moreover, when there is detuning, \ie, $\omega_0 \ne \omega_1$, Adler's
equation can also be used to estimate the locking range.

If we further assume that the waveform of each oscillator is also sinusoidal,
we can derive the Kuramoto model for coupled oscillators
\cite{acebron2005kuramoto} from the Adler's equation:
\begin{equation}\eqnlabel{Kuramoto1}
	\frac{d}{dt} \phi_i(t) = \Delta\omega_i + \omega_0A_c\sum_j J_{ij} \cdot \sin(\phi_i(t) - \phi_j(t)),
\end{equation}
where $\phi_i(t)$ is the phase of the $i$\textsuperscript{th} oscillator in the
system; $\omega_0$ is the frequency of the synchronized coupled oscillator
system; $\Delta\omega_i$ is the phase difference between each oscillator's
natural frequency and $\omega_0$; $A_c$ is a scalar representing the coupling
strength.

If we assume no frequency variation in all the oscillators for the moment, \ie,
$\Delta\omega_i=0$, and normalize $\omega_0$ to $1$, we can simplify
\eqnref{Kuramoto1} as
\begin{equation}\eqnlabel{Kuramoto2}
	\frac{d}{dt} \phi_i(t) = A_c\sum_j J_{ij} \cdot \sin(\phi_i(t) - \phi_j(t)).
\end{equation}

Similar analysis can be applied when we include SHIL in the system.
For a single oscillator, when its input has a periodic entry at $2\omega_1$, we
assume that the corresponding entry in $\vec v(t)$ also has a second harmonic
at $2\omega_0$, \ie, $\vec v_{(2\pi)}(t) = A_{v2}\sin(2t)$ and $\vec
u_{(2\pi)}(t) = A_{u2}\cos(2t + \Delta \phi_u)$,
Then following the same procedures as above, we get
\begin{equation}\eqnlabel{Adler2}
\frac{d}{dt} \Delta \phi(t) = \omega_0-\omega_1+\omega_0 A_2 \sin\left(2\Delta \phi(t) - \Delta \phi_u)\right),
\end{equation}
where $A_2=\frac{1}{2}A_{v2}A_{u2}$.
Steady state analysis of \eqnref{Adler2} indicates that the oscillator can lock
to the input with two stable phases, separated by $\pi$.

Incorporating \eqnref{Adler2} into the Kuramoto model \eqnref{Kuramoto2},
assuming that each oscillator is receiving an identical second-harmonic
periodic input SYNC, we can write the new coupled oscillator system equation as
\begin{equation}\eqnlabel{Kuramoto3}
	\frac{d}{dt} \phi_i(t) = A_c\sum_j J_{ij} \cdot \sin(\phi_i(t) - \phi_j(t))
						- A_s \cdot \sin(2\phi_i(t)),
\end{equation}
where scaler $A_s$ models the coupling strength from SYNC.

Note that the assumptions of sinusoidal PPV and sinusoidal waveforms are not
necessary.
Generalized Adler's Equation (GAE) \cite{BhRoASPDAC2009} has been developed for
handling arbitrary PPV shapes, and it provides good approximations to the
injection-locked solutions of \eqnref{PPVphi}.
This indicates that there are also generalized Kuramoto models where the $\sin()$
function can be replaced with arbitrary periodic functions, and oscillators can
be engineered to have the desired properties.

The Kuramoto model is a gradient system \cite{smale1961gradient}. 
There exists a global Lyapunov function for \eqnref{Kuramoto2}, which can be
considered as its ``energy'':
\begin{equation}\eqnlabel{E1}
	E(t) = A_c\sum_{i, j} J_{ij} \cdot \cos(\phi_i(t) - \phi_j(t)),
\end{equation}
such that
\begin{align}
	\frac{dE(t)}{dt} &= A_c\sum_{i} \left[ \frac{dE}{d\phi_i(t)} \cdot
                     \frac{d\phi_i(t)}{dt} \right] \\
				  &= A_c\sum_{i} \left[ - \left( \sum_{j} J_{ij} \cdot
					 \sin(\phi_i(t) - \phi_j(t))\right) \cdot \frac{d\phi_i(t)}{dt}
                     \right] \\
	              &= - \left|\left|\frac{d\vec \phi(t)}{dt}\right|\right|_2^2 \leq 0.
\end{align}

Therefore, the coupled oscillator system always attempts to minimize this
energy $E$.

$E$ as defined in \eqnref{E1} shares some similarities with the Ising Hamiltonian
in \eqnref{IsingH}.
If every oscillator's phase settles to a binary value of either $0$ or $\pi$,
corresponding to $s_i$ equal to $1$ or $-1$ in \eqnref{IsingH},
we have $\cos(\phi_i-\phi_j) = s_i \cdot s_j$.
Therefore, the coupled oscillators are naturally minimizing the Ising
Hamiltonian defined in \eqnref{IsingH}.

This reasoning is only valid when we assume that the phases of all the
oscillators settle to binary values.
It is easy to prove that the assumption holds for two oscillators --- they will
settle with a phase difference of either $0$ or $\pi$ depending on the polarity
of coupling coefficient $J_{12}$.
But as the number of oscillators increases, the analysis quickly becomes
difficult and the phases can indeed settle to non-binary values.
In other words, the system becomes an analog computer like oscillator-based
associative memory arrays rather than a digital combinatorial optimizer.

As we discuss in \secref{intro}, one key technique behind oscillator-based
Ising machines is the use of a second-harmonic input SYNC to make oscillators
behave like binary latches through the mechanism of SHIL.
This modification changes the Kuramoto model to \eqnref{Kuramoto3} and results
in a new Lyapunov function as follows.

\begin{equation}\eqnlabel{E2}
	E(t) = A_c\sum_{i, j} J_{ij} \cdot \cos(\phi_i(t) - \phi_j(t))
	  - \sum_{i} \frac{A_s}{2} \cdot \cos\left(2\cdot (\phi_i(t))\right).
\end{equation}

When SYNC is large enough, it enforces the phases $\phi_i$ to settle at either
$0$ or $\pi$, in which case the Lyapunov function can be simplified as
\begin{equation}\eqnlabel{E3}
	E(t) \approx A_c\sum_{i, j} J_{ij} \cdot \cos(\phi_i(t) - \phi_j(t))
	  - \frac{1}{2}(n+1)\cdot A_s,
\end{equation}
where $n+1$ is the total number of oscillators, with the last one being the
phase reference always representing logic 1.

Therefore, the introduction of SYNC does not change the relative ``energy''
levels between valid binary phase configurations; it modifies them by the same
amount.
It does not change the location of the ground state.

\ignore{
	Steady states of such a coupled oscillator network can be reached by
	several types of ``annealing'' --- decreasing noise, increasing coupling, and
	pumping up SYNC.
	It is provable that every steady state corresponds to a local optimum of the
	Ising Hamiltonian.
	In the presence of noise, with sufficient annealing time, the network will tend
	to settle to the most stable state, corresponding to the ground state of Ising
	Hamiltonian, \aka{} the global optimum of the combinatorial problem.

	SHIL makes almost any oscillator work as a logic latch that can store one bit
	of information; storing this bit securely enables oscillator-based Boolean
	computation. An oscillator latch is also very useful even when the bit is not
	entirely secure. Due to phase noise and detuning in the oscillator, the
	phase-based bit stored in the latch is a random process --- it flips from now
	and then. Everytime it is measured, it can randomly assume 0 or 1 value. An
	oscillator becomes a binary random number generator. Given today's oscillator
	technology, they can be very compact compared with existing solutions.

	Moreover, the probability of this random variable at a given time can be
	controlled through coupling between oscillator latches. For example, if there
	are two oscillators under weak SHIL, their phase values can represent {0, 0},
	... {1, 1} (four sets of values). If they are connected through a resistor (in
	electrical domain), such that the phase of one of them affects the other. As is
	obvious to see, if the coupling is strong enough, I.e., the resistor is very
	small such that the two oscillators are connected to each other through a short
	connection, their phases can only settle to {0,0} and {1,1} values. Similarly,
	when the coupling is chosen appropriately, {0,1} and {1,0} may still have some
	probability, but much lower than those of the other "ground states". If we let
	this two oscillator system settle to the steady state, their phases will have
	solved an optimization problem (or a logic problem) whose solutions require that
	the two values are the same. More complex problems can be represented and
	solved with more complicated coupling between the oscillator latches.

	Recently, Ising machines have been proposed for invertible logic computation
	applications, such as factorization. The idea is similar. If the logic problem
	is formulated as an optimization problem such that the ground states correspond
	to its valid solutions, when the Ising model settle to its ground state, the
	solutions solve the logic problem. In this way, a connection for addition can
	also be used to solve for subtraction; multipliers can be used for
	factorization.

	Compared with the existing ideas on the practical implementation of Ising
	machines, our idea of oscillator-based Ising machine has several advantages
	(features). Firstly, the oscillator technology is ready available, and general
	across many fields of engineering. The idea can be demonstrated (or
	experimented with) using well-developed CMOS oscillators. It can also be
	implemented with novel oscillator devices, such as spin torque, MEMS,
	memristor-based ones. Many of them are nano-scale and energy efficient. Many
	are also large-scale integrable  on chip already. Note that on-chip Ising
	machine has been proposed and implemented before, with standard CMOS
	level-based latches. But applications such as general combinatorial
	optimization or invertible logic have not been achieved; the only demonstration
	was pattern recognition, the standard task suitable for hopfield network (which
	is deterministic, no need to control noise, easy to control coupling).

	Oscillator-based Ising machines normally require sparse connections between
	vertices. This is rarely a problem since most real world problems can be
	formulated into sparse coupling [], or even into a grid of nodes [] if
	auxiliary nodes are introduced.
	Recent work [] use FPGA to form virtual connections, in order to achieve full
	connection among vertices.
	The same idea still applies to oscillator-based Ising machines if desired.

	We shall not confuse oscillator-based Ising machines with quantum annealers,
	such as those being developed at D-Wave [].
	No quantum annealing is available.
	The performance will be worse for larger-scale problems.
	But since oscillators are easily integrable and they mostly work at room
	temperature, they are cost and energy efficient, and can scale faster and more
	easily.
	Also, more auxiliary vertices normally result in better convergence to the
	ground state of the original optimization problem. 
}

One major obstacle to the practical implementation of large-scale Ising
machines is variability.
Researchers of SRAM-based CMOS Ising machines explicitly attribute the
``limited efficacy'' to the variations in SRAMs \cite{yamaoka2016IsingCMOS}.
Indeed, SRAMs are not good random number generators --- process variations
often give them preferences of generating either 1s or 0s.
Ising machines based on nano-magnets \cite{camsari2017pbits} are likely to
suffer from the same problem in hardware implementations.
Specifically, each nano-magnet will prefer either the ``up'' or ``down'' state
after fabrication, the effects of which are yet to be studied.
In comparison, the binary states in an oscillator are defined based on
phase/timing, and are thus perfectly ``symmetric'' --- there is no mechanism
making an oscillator prefer phase 0 to phase $\pi$ under SHIL.
This symmetry can lead to markedly improved performances over the SRAM-based
scheme in large-scale implementations.

Even so, there are still variations in coupled oscillators, coming from another
source --- the variability of the oscillators' natural frequencies.
Taking this into consideration, we rewrite the Kuramoto model in
\eqnref{Kuramoto3} as
%
\begin{equation}\eqnlabel{Kuramoto_var}
	\frac{d}{dt} \phi_i(t) = \Delta\omega_i
	                    + A_c\sum_j J_{ij} \cdot \sin(\phi_i(t) - \phi_j(t))
						- A_s \cdot \sin\left(2\phi_i(t)\right).
\end{equation}

And the corresponding Lyapunov function becomes
\begin{equation}\eqnlabel{E4}
	E(t) = \sum_{i} \Delta\omega_{i} \cdot \phi_i(t)
	    + A_c\sum_{i, j} J_{ij} \cdot \cos(\phi_i(t) - \phi_j(t))
	    - \frac{1}{2}(n+1)\cdot A_s.
\end{equation}

Note that \eqnref{E4} differs from \eqnref{E3} only by a weighted sum of
$\phi_i$ --- it represents essentially the same ``energy'' landscape but tilted
linearly with the optimization variables.
While it can still change the locations and values of solutions, its effects
are easy to analyze given a specific combinatorial optimization problem.
Also, as the coupling coefficient $A_c$ gets larger, the effect of detuning is
reduced.

The OPO-based Ising machine also uses phase-based logic encoding
\cite{marandi2014network}, but it does not use self-sustaining oscillators.
The variations in frequency and amplitude are minimized by generating the
pulses from the same laser and letting them travel through the same optical
fiber.
As such, the implementation is not easy to miniaturize.
One way towards integration is to use individual ring resonators as parametric
oscillators.
Then process variations in integrated photonics will create problems again.
And analyzing variation's effects on parametric oscillators appears to be an
unsolved problem; it is much more difficult than the analysis above for coupled
oscillators.

Our analysis so far focuses only on the deterministic model of oscillator-based
Ising machines, thus is not complete.
Starting from a random initial condition, coupled oscillators evolve in a
deterministic manner, in an effort to minimize the global Lyapunov function $E$
in \eqnref{E3}, which corresponds to the Ising Hamiltonian in \eqnref{IsingH}.
This is a dynamical-system-based implementation of the gradient descent
algorithm that can find local minima of a function.
Therefore, like gradient descent, its effectiveness in finding the global
minimum is limited.
However, we can reasonably suspect that, if there is noise in the system, it
becomes more likely for the coupled oscillators to settle to the minima with
lower $E$s; like simulated annealing, the probability of achieving the ground
state becomes higher.

To analyze this hypophysis, we first introduce noise into the coupled
oscillator model.
One common way is to assume there is white noise in the central frequencies,
which can be modelled as additive white noise in the Kuramoto model.
\begin{equation}\eqnlabel{noise1}
	\frac{d}{dt} \phi_i(t) = A_c\sum_j J_{ij} \cdot \sin(\phi_i(t) - \phi_j(t))
						- A_s \cdot \sin(2\phi_i(t)) + A_n\xi_i(t),
\end{equation}
where $\xi_i(t)$ represents Gaussian white noise with zero mean and
correlator $\langle\xi_i(t),~\xi_i(\tau)\rangle = \delta(t-\tau)$;
scaler $A_n$ represents the magnitude of noise.

\eqnref{noise1} can be rewritten as a stochastic differential equation (SDE).
\begin{equation}\eqnlabel{noise2}
	d\phi_{it} = \left[A_c\sum_j J_{ij} \cdot \sin(\phi_{it} - \phi_{jt})
						- A_s \cdot \sin(2\phi_{it}) \right]dt + A_ndW_t.
\end{equation}

From it, one can derive master equations describing the time evolution of the
probability of the system to occupy each state.
Since we are mainly interested in the steady state, here we can directly apply
the Boltzmann law from statistical mechanics \cite{landau1969statistical}.
For a system with discrete states $\vec s_i$, $i=1,\cdots,M$, if each state is
associated with an energy $E_i$, the probability $P_i$ for the system to be at
each state can be written as follows.
\begin{equation}
	P_i = {\frac{e^{- E_i / k T}}{\sum_{j=1}^{M}{e^{- E_j / k T}}}},
\end{equation}
where $k$ is the Boltzmann constant, $T$ is the thermodynamic temperature of
the system.
While $k$ and $T$ are concepts specific to statistical mechanics, in this
context the product $kT$ correlates with the magnitude of $A_n$.

Given two spin configurations $\vec s_1$ and $\vec s_2$, the ratio between
their probabilities is known as the Boltzmann factor:
\begin{equation}
\frac{P_2}{P_1} = e^{\frac{E_1 - E_2}{kT}}.
\end{equation}

In oscillator-based Ising machines, the energy at a spin configuration is
\begin{align}
	E(\vec s) &= A_c\sum_{i, j} J_{ij} s_is_j - \frac{1}{2}(n+1)\cdot A_s \\
	     &= A_c\sum_{i, j} J_{ij} s_is_j - Const, \eqnlabel{E5}
\end{align}

Therefore,
\begin{equation}
E_1 - E_2 \propto A_c.
\end{equation}

If $\vec s_1$ is the higher energy state, \ie, $E_1 > E_2$, as the coupling
strength $A_c$ increases, it becomes less and less likely for the system to
stay at $\vec s_1$. The system prefers the lowest energy state in the presence
of noise.

Note that the Boltzmann law describes systems with states that have physical
energies.
A coupled oscillator system, like many computational systems, its dissipative
in its nature, \ie, it is a thermodynamically open system operating far from
equilibrium.
As such, it does not have a physical energy associated with its states.
However, it is provable that a global Lyapunov function, if it exists, can be
used as an energy function to derive the same Boltzmann law
\cite{yuan2010lyapunov}.
Therefore, the above reasoning for achieving better minima under noise still
holds for oscillator-based Ising machines.

The operation of oscillator-based Ising machines modelled in \eqnref{noise1} is
controlled by several parameters --- the mutual coupling strength $A_c$, the
SYNC coupling strength $A_s$, and the noise level $A_n$.
Because of the Boltzmann law, we normally ramp up $A_c$ slowly in order to keep
the system at the thermodynamic equilibrium all the time.
$A_s$ can be kept constant or ramped up in the meanwhile.
In fact, $A_c$, $A_s$, $A_n$ can all be time varying, resulting in various
annealing profiles.
As we show in \secref{examples}, this property gives us much flexibility in the
engineering of oscillator-based Ising machines.

\section{Speed and Scalability}\seclabel{speed}

While examining the mechanism for coupled oscillators to minimize the Ising
Hamiltonian, we have noted the similarity between oscillator-based Ising
machines and conventional algorithms such as gradient descent and simulated
annealing.
But unlike these algorithms run on conventional computers, Ising machines
compute in a highly parallel fashion, and are widely believed to have better
scalability for large-sized problems.
The scaling of simulated annealing's runtime depends largely on the
combinatorial optimization problems.
But in the case of Ising machine, the computation time remains mostly constant
from our observation, with the hardware size growing linearly or quadratically
depending on the problems \cite{lucas2013ising};

The computation time is essentially the convergence rate of a coupled
oscillator network.
Existing study on this subject is rather scattered
\cite{jadbabaie2004kuramoto,wang2013expsync,dorfler2014synchronization,patra2016lyapunov,coletta2017finite}.
Exact analytical solutions of the convergence rate of the Kuramoto model are
difficult to acquire \cite{jadbabaie2004kuramoto}.
In some studies, the spectrum of Lyapunov exponents of the locked states in the
Kuramoto model are used to analyze its speed and calculated for specific
problems \cite{patra2016lyapunov,coletta2017finite}; how it scales with the
problem size is yet to be studied.

Therefore, in this section, we analyze the convergence speed of coupled
oscillators through a computational study.
To do so, several factors need to be taken into consideration.
Does the convergence rate depend on the network's size?
Does it depend on the sparsity?
Does it depend on the type of connections?
To answer these questions, we need to run simulations on coupled oscillator
networks with various configurations.

Firstly, we simulate fully connected networks of different sizes, with random
connection weights uniformly distributed between $0$ and $1$, starting from
random initial conditions uniformly distributed between $0$ and $\pi$.
At each size, 10 samples are simulated.
We visualize the convergence rate by plotting the energy function, which
monotonically decreases with time.
From \figref{plot_Kuramoto_full_binary_speed}, we observe that networks with
different numbers of oscillators converge at approximately the same speed.

\begin{figure}[htbp!]
    \centering
    {
        \epsfig{file=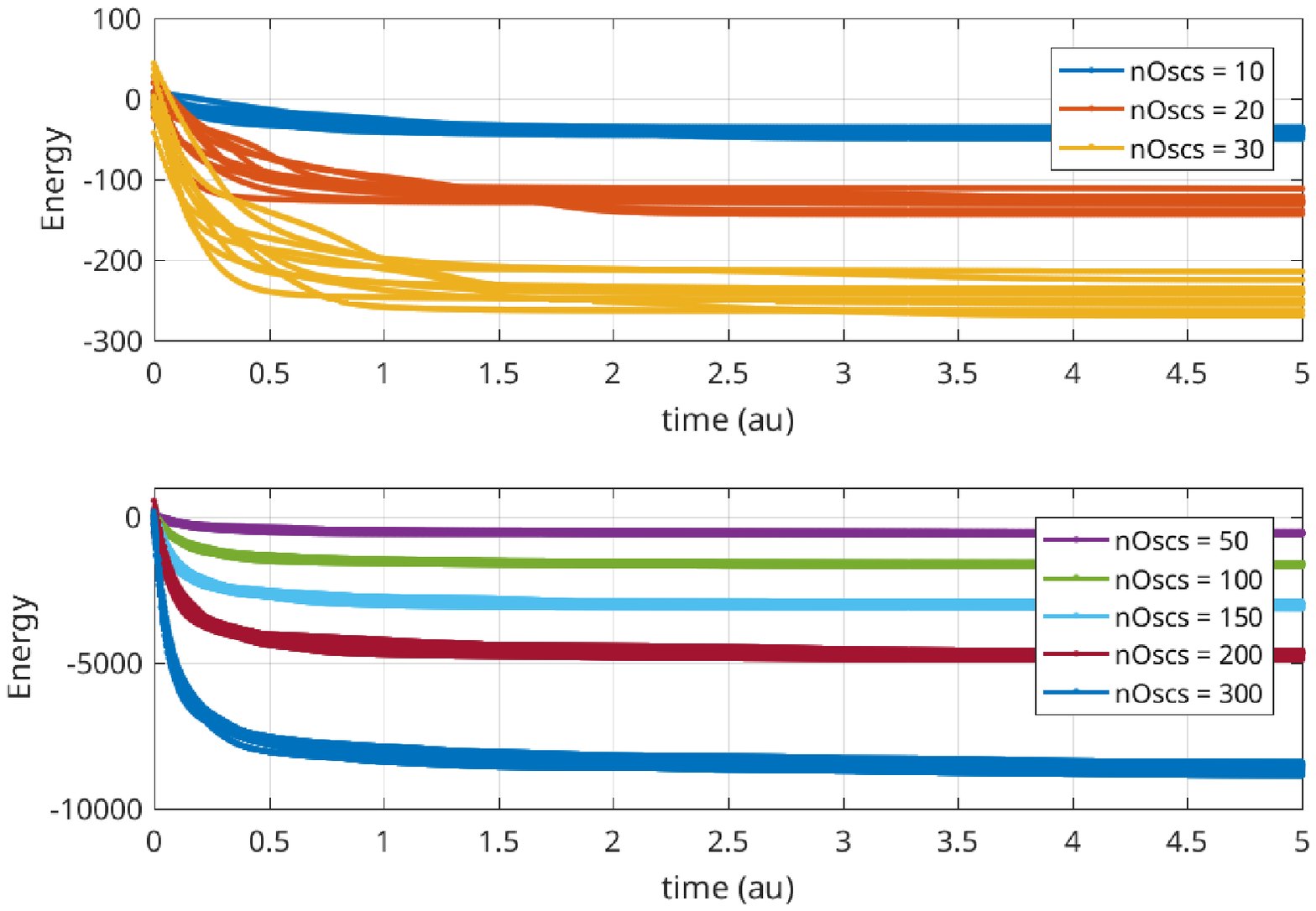, width=0.9\linewidth}
    }
    \caption{Energy \vs time for fully connected oscillator networks.
    \figlabel{plot_Kuramoto_full_binary_speed}}
\end{figure}

Furthermore, we fix the size of the network to 100 and vary the sparsity of the
connections.
The sparsity is defined as the ratio between the actual number of non-zero
connections and the possible number of connections in a full network.
Again, with each sparsity, 10 random samples are simulated; we plot the energy
functions in \figref{plot_Kuramoto_sparse_binary_speed}.
From it, we see that with a fixed size, as the network gets sparser, the
objective energy becomes shallower, and the convergence rate decreases
marginally.

\begin{figure}[htbp!]
    \centering
    {
        \epsfig{file=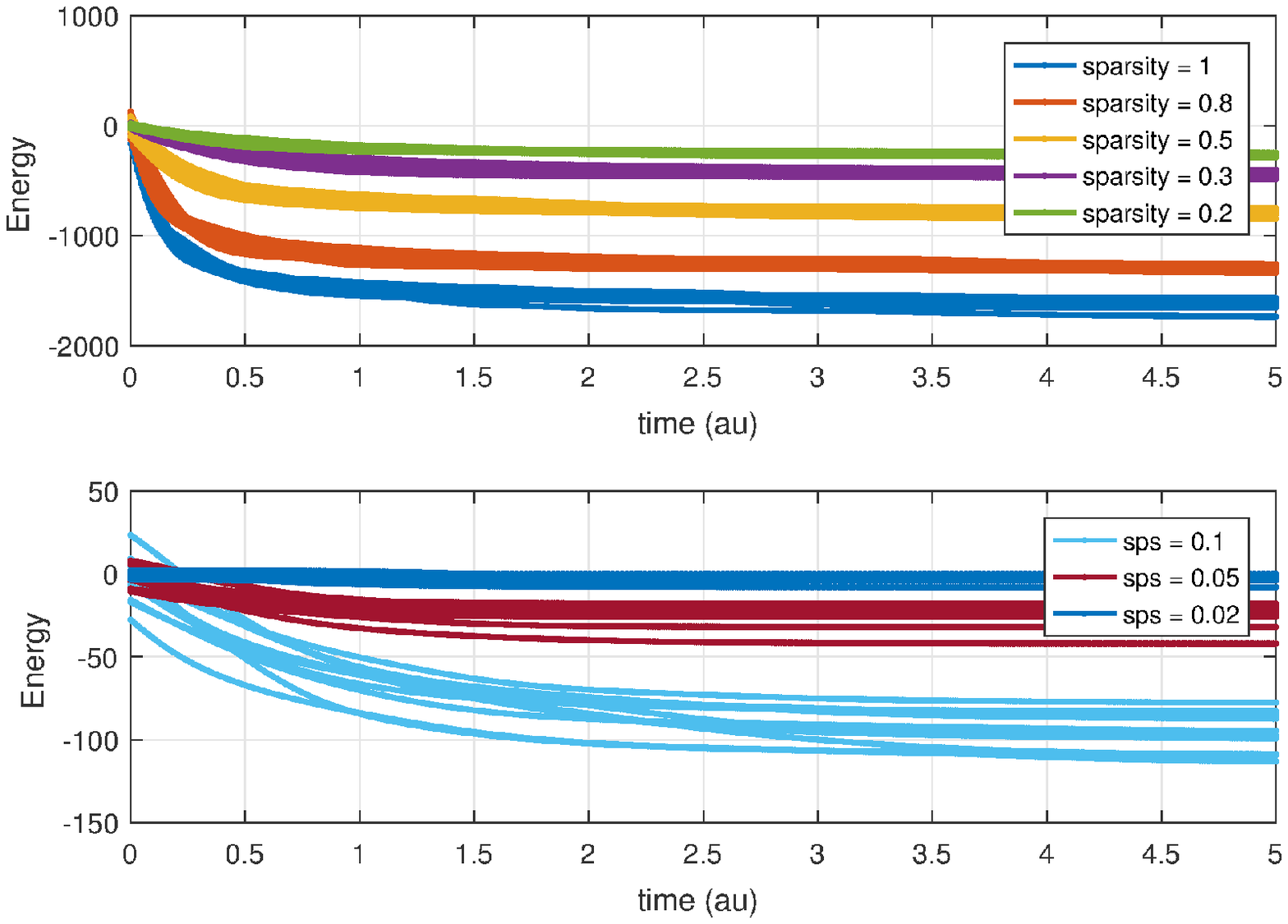, width=0.9\linewidth}
    }
    \caption{Energy \vs time for sparse oscillator networks.
    \figlabel{plot_Kuramoto_sparse_binary_speed}}
\end{figure}

In a fully connected network, both the average and largest distances between
oscillators are $1$.
In sparse random graphs, the largest distance, \aka, the diameter, increases
very marginally\footnote{For a graph with $n$ vertices and a connection
probability of $p$, the diameter is in the order of $\frac{\log(n)}{\log(np)}$
\cite{chung2001diameter}, which grows slower than $\log(n)$ given any $p$.}
as the network becomes sparser \cite{chung2001diameter}.
Since the convergence rate in both cases does not change much with size or
sparsity, it is natural to suspect that it may instead scale with the average
or maximum distance between oscillators in the network.
To study this possibility, we conduct further experiments and simulate the
``worse case'' for connection distance --- all oscillators coupled are in a
single line.
In this case, both the average and largest distances grow linearly with the
number of oscillators.
From the energy functions plotted in \figref{plot_Kuramoto_binaryline_speed},
we observe that even in this connection configuration, the speed does not
change much with increased network size.
These results are encouraging. As the problem size grows, the hardware size
does need to increase \cite{lucas2013ising}, but the computation time remains
almost constant.

\begin{figure}[htbp!]
    \centering
    {
        \epsfig{file=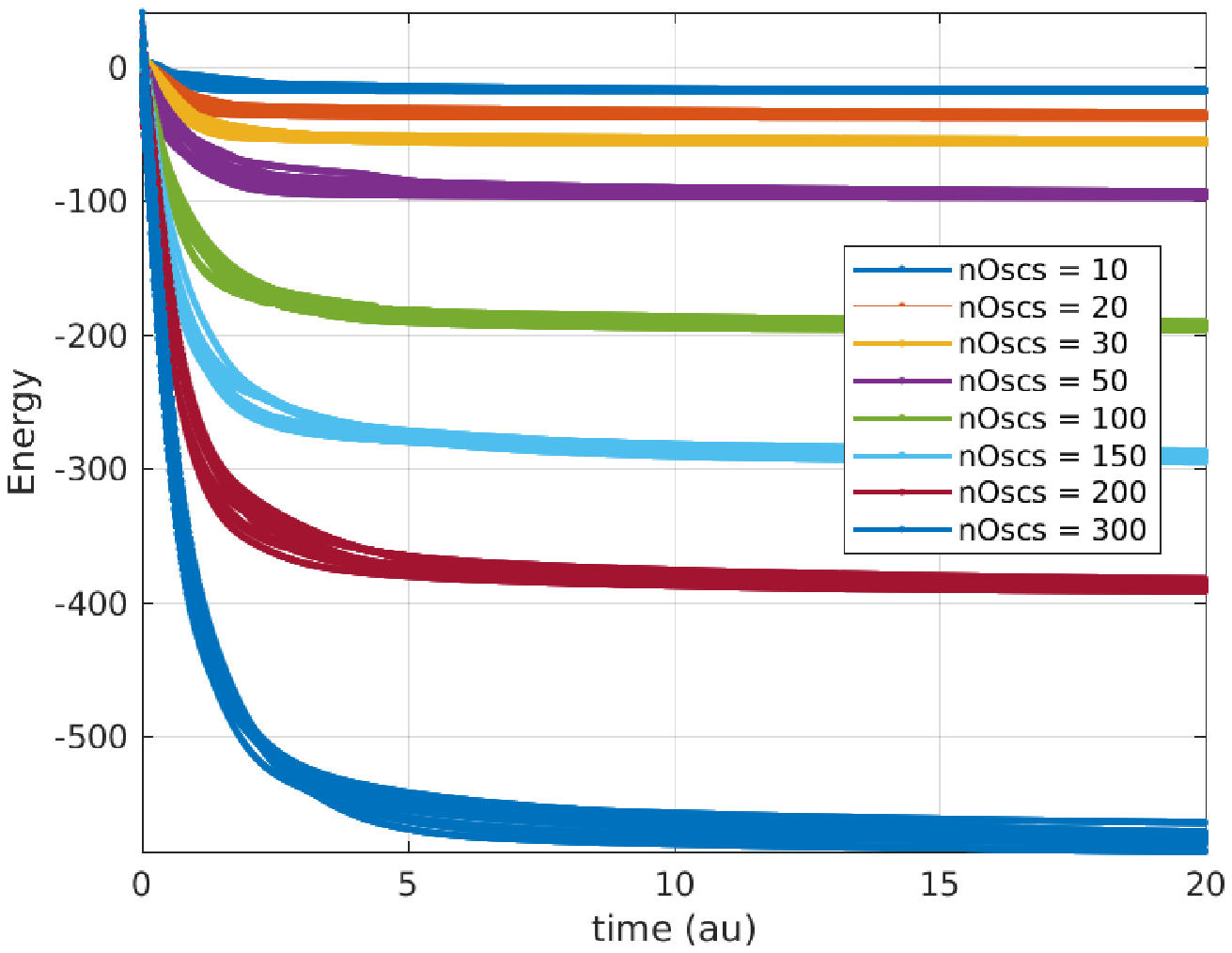, width=0.9\linewidth}
    }
	\caption{Energy \vs time for oscillators coupled in a single line.
    \figlabel{plot_Kuramoto_binaryline_speed}}
\end{figure}

Note that in all these results from the Kuramoto model, time is measured in
seconds.
The results are based on the Kuramoto model defined \eqnref{Kuramoto3}, where
we assume $\omega_0 = 1$.
In fact, from \eqnref{Kuramoto1} we can see that the run time is inversely
proportional to the product of $\omega_0$ and $A_c$.
This is to say, if the oscillator's frequency is at GHz scale, the time to
synchronize becomes nanoseconds as opposed to seconds.
Of course, for nano-oscillators, the coupling strength $A_c$ is normally not
uniformly distributed between 0 and 1, as we have assumed in the results in
this section. 
$A_c$ controls the number of cycles for the oscillators to synchronize; its
value depends not only on the resistive coupling between oscillators, but also
on the type of oscillator.
As we show in \secref{examples}, it may take LC oscillators coupled with
$M\Omega$ resistors 100 cycles to synchronize their phases.
Even so, it takes only a fraction of a microsecond for computation.
While the speed is already appealing, employing other oscillator technologies
can further improve it.
Ring oscillators in standard CMOS technologies can nowadays achieve 100GHz, and
are known to injection lock and synchronize quickly.
Resonant Body Transistors (RBTs), a type of silicon-based CMOS-compatible MEMS
resonators, have been demonstrated to achieve $>$10GHz frequency
\cite{weinstein2010RBT}.
Spin-torque oscillators operate at frequencies of tens of GHz
\cite{Pufall:2005:STNO_FM:2005ApPhL..86h2506P}, offering exciting power and
speed possibilities.

\section{Examples}\seclabel{examples}

In this section, we demonstrate the feasibility of oscillator-based Ising
machines with several examples.

\subsection{Small MAX-CUT Problems}\seclabel{smallMAXCUT}

The MAX-CUT problem is a combinatorial optimization problem related to a graph,
where we try to find a subset of vertices such that the total weights of the
cut set between this subset and the remaining vertices are maximized.
The MAX-CUT problem is one of Karp's 21 NP-complete problems \cite{karp1972np}.
They have a direct mapping to the Ising model in \eqnref{IsingH}, with $J_{ij}
= J_{ji}$ representing the weight between node $i$ and node $j$.
Then we can write the relationship between the Hamiltonian function and the cut
size as
\begin{equation}
	H = \sum_{i, j} J_{ij} s_i s_j = \sum_{i, j} J_{ij} - 2 S_c,
\end{equation}
where $S_c$ is the cut size --- it is maximized when the Ising Hamiltonian is
minimized.

We construct a small-sized MAX-CUT problem with a full graph of six vertices
using the Kuramoto model in \eqnref{Kuramoto3}.
Each edge has a random weight drawn from a uniform distribution between 0 and
2.
The magnitude of SYNC is fixed at $A_s=2$, while we ramp up the coupling
strength $A_c$ from 0 to 5.
Results from the deterministic model ($A_n=0$) and the stochastic model
($A_n=0.1$) are shown in \figref{MAXCUT6} and \figref{MAXCUT6-noise}
respectively.
In the plots, oscillators are started with random phases between 0 and $\pi$;
after a while, they all settle to one of the two phase-locked states separated
by $\pi$.
These two groups of oscillators represent the two subsets of vertices in the
solution.
In this case, the results reliably return \{2,3,6\} and \{1,4,5\}.
For this size-6 problem, we have enumerated all the possible cut sets --- the
result from phase-based simulation is indeed the global optimal solution.

\begin{figure}[htbp]
	\begin{minipage}{0.49\linewidth}
      \epsfig{file=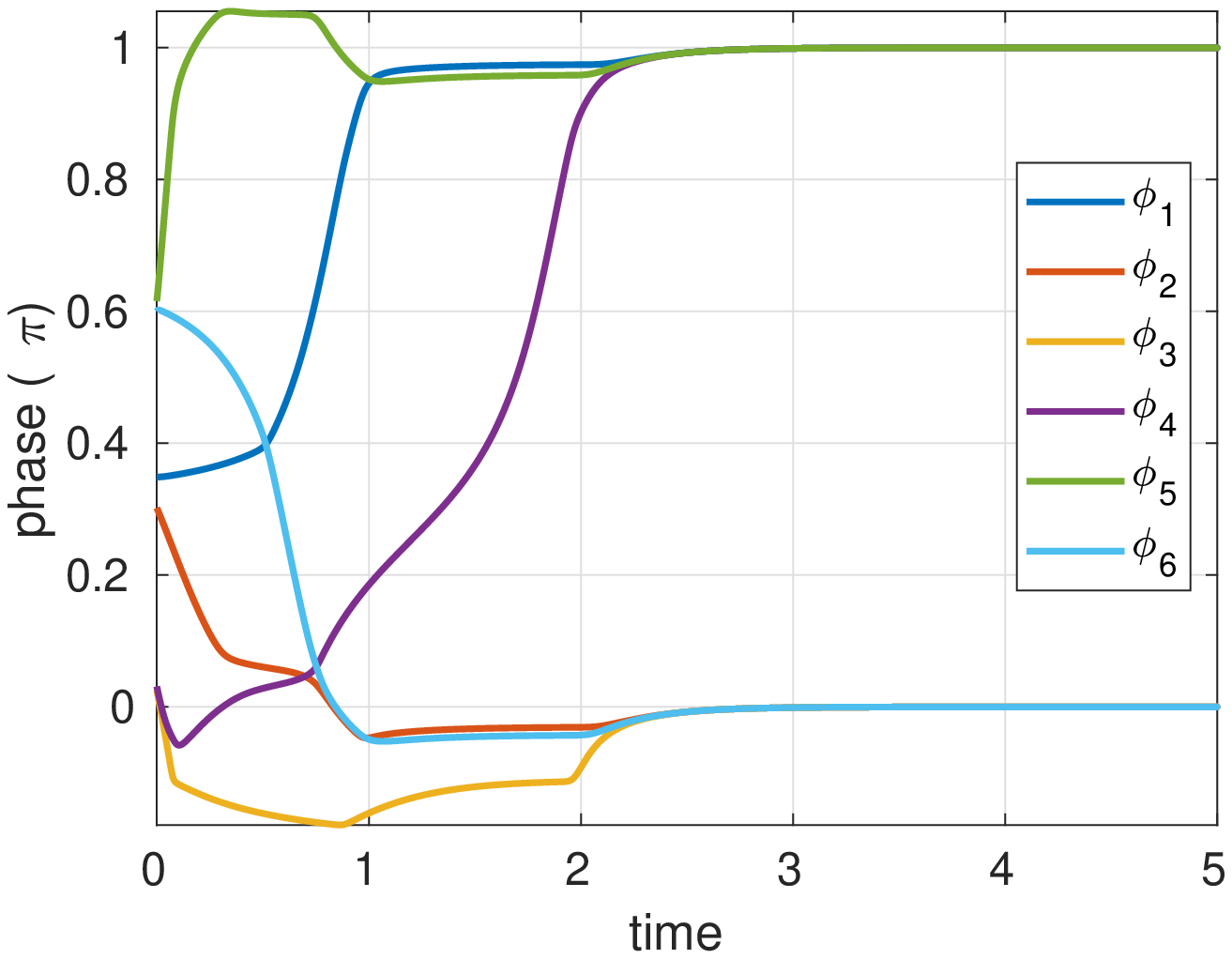,width=\linewidth}
	  \caption{Phases of oscillators solving a size-6 MAX-CUT problem without noise.}\figlabel{MAXCUT6}
	\end{minipage}
    \hfill
	\begin{minipage}{0.49\linewidth}
      \epsfig{file=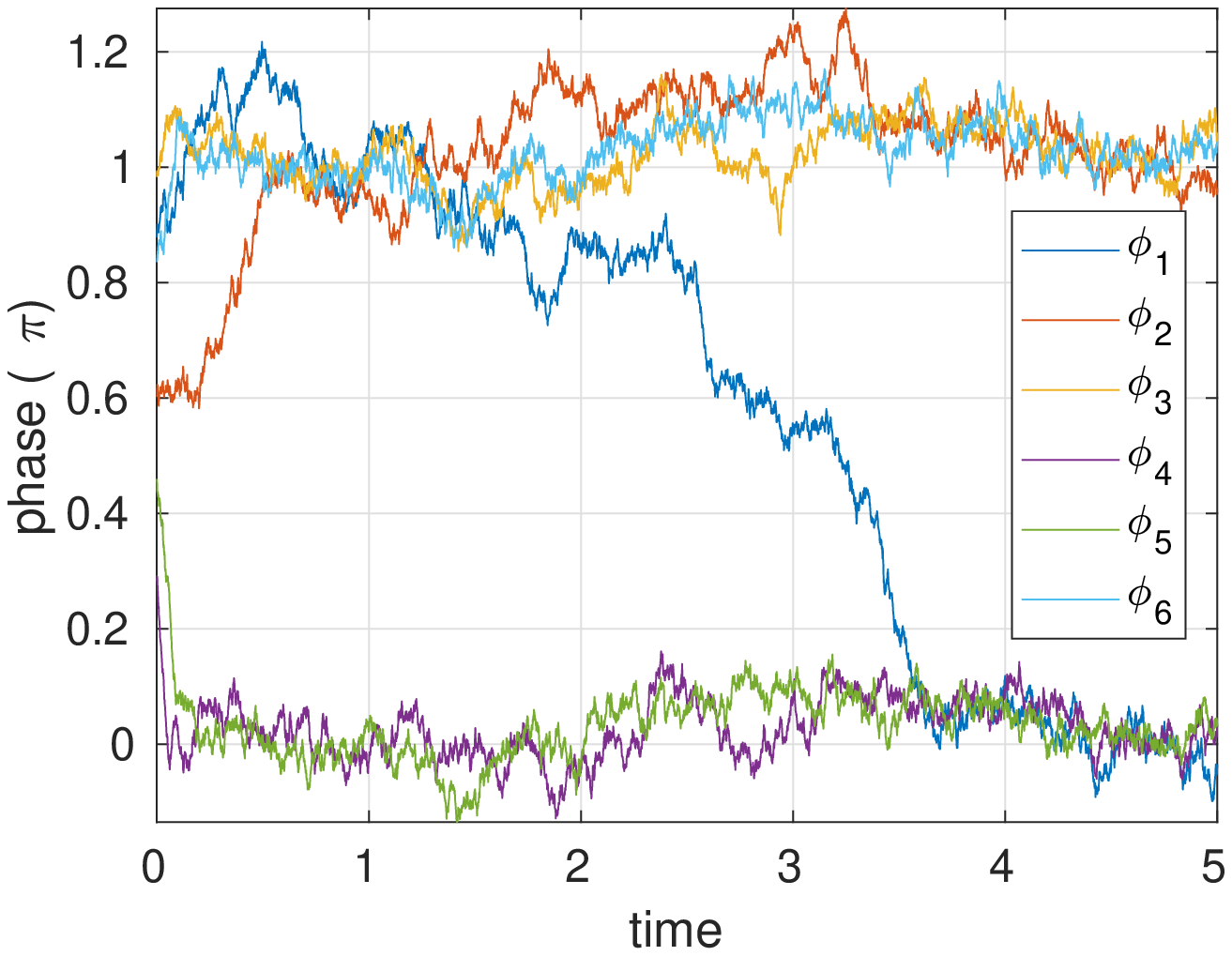,width=\linewidth}
	  \caption{Phases of oscillators solving a size-6 MAX-CUT problem with noise.}\figlabel{MAXCUT6-noise}
	\end{minipage}
\end{figure}

The Kuramoto models are prototyped in a \MATLAB-based simulation platform MAPP
\cite{WaAaWuYaRoCICC2015MAPP,WaKaWuRo2015MAPPnemo}.
Here in the paper, we show the minimum code for generating the same results in
Listings \ref{lst:KuramotoF_sin} and \ref{lst:run_MAXCUT_6}.
Note that these are SDE simulations with random initial conditions.
Every run will return different waveforms; there is no guarantee for achieving
the ground state every time.

\matlabscript{code/KuramotoF_sin.m}{lst:KuramotoF_sin}{\texttt{KuramotoF\_sin.m}}
\matlabscript{code/run_MAXCUT_6.m}{lst:run_MAXCUT_6}{\texttt{run\_MAXCUT\_6.m}}

Instead of using phase macromodels, we can also directly simulate oscillators'
DAEs as in \eqnref{DAE} and achieve the same results.
Such simulations are at a lower lever than macromodels and less efficient.
But they are closer to physical reality and are standard in circuit design.
In the simulations, six cross-coupled LC oscillators are tuned to a frequency
of 1GHz.
Each pair of oscillators $i$ and $j$ are coupled through a resistor with
conductance $G_0 \cdot J_{ij}$ between the opposite differential nodes,
where $G_0 = 1/1M\Omega$.
In this way, a positive resistor tends to develop opposite phases in the two
oscillators, same as the effect of a positive $J_{ij}$ in the Ising
Hamiltonian.
Results from transient simulation using ngspice-26 are shown in
\figref{MAXCUT_LC_6}.
The six oscillators' phases settle into the correct two groups \{2,3,6\} and
\{1,4,5\} within 0.1$\mu$s, which is about 100 cycles of oscillation.
We have tried this experiment with different sets of random weights, starting
from different random initial conditions; SPICE simulations on oscillators'
DAEs reliably return the optima of various size-6 MAX-CUT problems.

\begin{figure}[htbp!]
    \centering
    {
        \epsfig{file=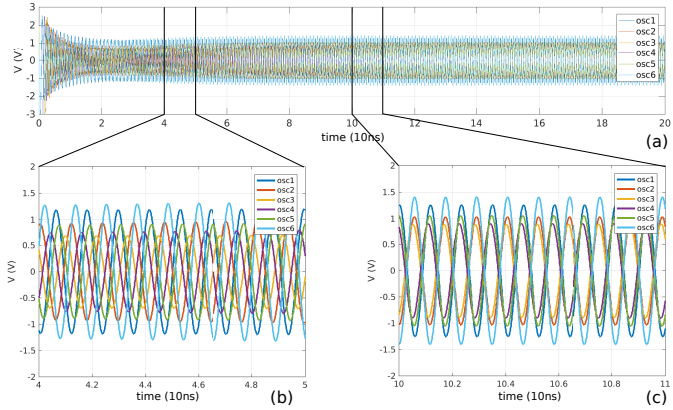, width=\linewidth}
    }
    \caption{Simulation results from ngspice on 6 coupled oscillators.
    \figlabel{MAXCUT_LC_6}}
\end{figure}

On a breadboard, we implement an even smaller MAX-CUT problem by coupling four
LC oscillators together.
The CMOS devices are implemented with chips ALD1106/7.
The inductors are of size 10mH; capacitors are 68nF.
The central frequency is about 38kHz.
And the potentiometers are of maximum resistance of 220kOhm. By tweaking the six
potentiometers, we can conveniently adjust the edge weights to try various
size-4 MAX-CUT problems.
The results are observed using a four channel oscilloscope, as shown in
\figref{Ising-LC} (c).
Through experiments with various sets of weights, we have validated that this
is indeed a proof of concept hardware implementation of oscillator-based Ising
machines, the first of its kind.

\begin{figure}[htbp!]
    \centering
    {
        \epsfig{file=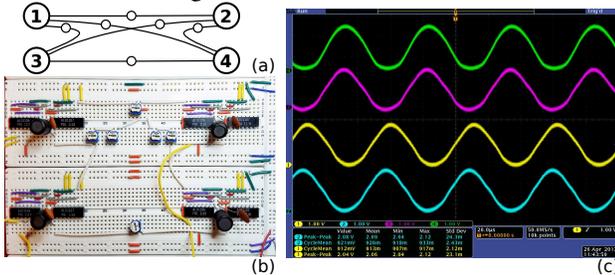, width=0.9\linewidth}
    }
    \caption{A simple oscillator-based Ising machine solving size-4 MAX-CUT
problems: (a) illustration of the 6 connections of 4 units; (b) breadboard
implementation with 4 CMOS LC oscillators and 6 potentiometers;
(c) oscilloscope measurements showing the maximum cut is between nodes \{1, 2\}
and \{3, 4\}.
    \figlabel{Ising-LC}}
\end{figure}

\subsection{A Larger MAX-CUT Problem}\seclabel{largerMAXCUT}

In this section, we test the proposed scheme on a larger-sized benchmark MAX-CUT
--- G22 from \cite{festa2002randomized}\footnote{G22 is available for download in
set1 at http://www.optsicom.es/maxcut.}.
The graph has 2000 vertices and 19990 edges.
We perform phase macromodel simulations with 2000 coupled oscillators, starting
from random initial phases.
In the process, we pump up the coupling strength $A_c$ while keeping the noise
level $A_n$ constant.
Instead of keeping the coupling from SYNC $A_s$ also constant, we find that
ramping it up and down improves the results.
\figref{MAXCUT_G22} Shows how the phases evolve over time.
The corresponding cut size is plotted in \figref{MAXCUT_G22_E}.
The code for generating the results is shown in Listing \ref{lst:KuramotoF} and
\ref{lst:run_MAXCUT_G22}.
We have run the experiments with the same parameters 100 times; the mean and
maximum cut sizes are shown in \tabref{results}.

\begin{figure}[htbp!]
    \centering
    {
        \epsfig{file=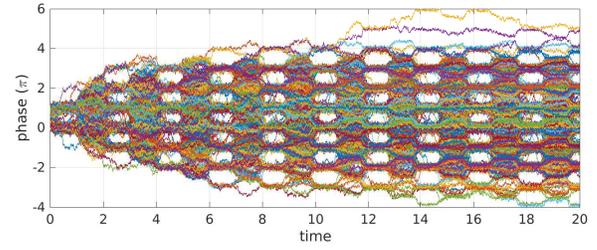, width=\linewidth}
    }
	\caption{Phases of oscillators solving the G22 MAX-CUT benchmark
             problem \cite{festa2002randomized}.
    \figlabel{MAXCUT_G22}}
\end{figure}

\begin{figure}[htbp!]
    \centering
    {
        \epsfig{file=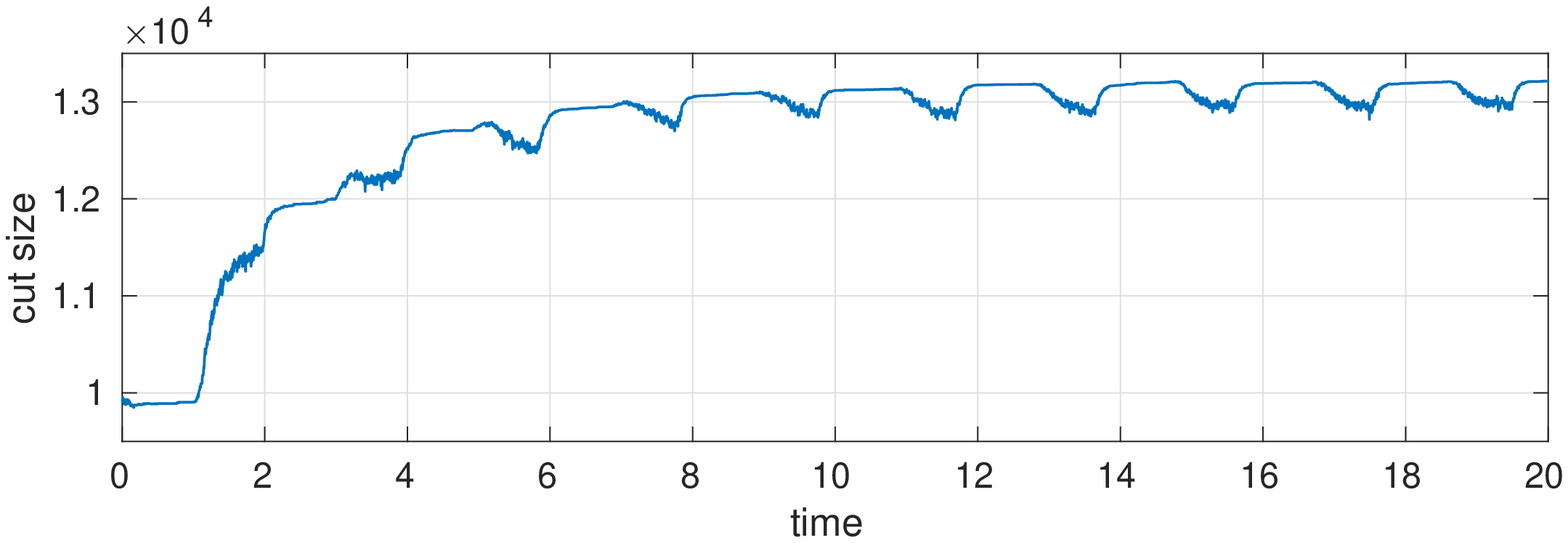, width=\linewidth}
    }
    \caption{Cut size \vs time corresponding to results in \figref{MAXCUT_G22}.
    \figlabel{MAXCUT_G22_E}}
\end{figure}

\matlabscript{code/KuramotoF.m}{lst:KuramotoF}{\texttt{KuramotoF.m}}
\matlabscript{code/run_MAXCUT_G22.m}{lst:run_MAXCUT_G22}{\texttt{run\_MAXCUT\_G22.m}}

\begin{table}[htb]
    \begin{center}
        \begin{tabular}{|m{0.3\linewidth}|m{0.2\linewidth}|m{0.2\linewidth}|}
        \hline
        {\bf ~~} &
        {\bf mean in 100} &
        {\bf best in 100} \\ \hline
        \hline
        CIM\cite{inagaki2016ScienceIsing2000} & 13248 & 13313
        \\ \hline
        our scheme & 13253 & 13305
        \\ \hline
        without noise & 13203 & 13249
        \\ \hline
        without SYNC & 13050 & 13155
        \\ \hline
        sinusoidal PPV & 13214 & 13276
        \\ \hline
        1\% var. in freq. & 13249 & 13309
        \\ \hline
        5\% var. in freq. & 13252 & 13303
        \\ \hline
        \end{tabular}
    \end{center}
  \caption{Results of oscillator-based Ising machines with different configurations 
		   run on MAX-CUT benchmark G22, compared with coherent Ising machine
           (CIM) \cite{inagaki2016ScienceIsing2000}. \tablabel{results}}
\end{table}

More experiments have been conducted on this benchmark to demonstrate the
mechanism of oscillator-based Ising machines.
We have tried removing noise from the model, \ie, $A_n=0$;
the solutions are considerably worse, as shown in \tabref{results}.
We have also tried this Ising machine without SYNC, \ie, changing \texttt{f2}
in Listing \ref{lst:run_MAXCUT_G22} to always return 0.
Then the coupled oscillators become the same as an associative memory array
people use for specialized image processing tasks
\cite{Hoppensteadt2000synchronization,maffezzoni2015oscArray,Porod2015physical}.
As shown in \tabref{results}, we have observed significantly worse results from
such coupled-oscillator-based associative memories.
SYNC and the mechanism of SHIL are indeed essential in the operation of
oscillator-based Ising machines.

Moreover, in Listing \ref{lst:KuramotoF}, we have changed the Kuramoto model,
making the $\sin()$ term in \eqnref{Kuramoto1} a smooth square function.
This changes the $\cos()$ term in the energy function \eqnref{E1} to a triangle
function.
Such a change appears to give better results than the original.
It requires designing oscillators with special PPVs and waveforms such that the
convolution of them is a square wave, \eg, oscillators with square PPVs and
spiky waveforms, or {\it vice versa}.
This is not difficult in practice.
In fact, rotary traveling wave oscillators naturally have square PPVs
\cite{chen2011RTWO}.
Ring oscillators can also be designed with various PPVs and waveforms by sizing
each stage individually.
The best phase macromodel with the optimal energy function for oscillator-based
Ising machines is yet to be studied. 
But whatever it may be, oscillators are versatile enough to be designed with
the desired properties.

We can also assume there is some variability in the central frequencies of
oscillators.
In Listing \ref{lst:KuramotoF}, we show results from simulating
equation \eqnref{Kuramoto_var}, with $\Delta\omega_i$ from a Gaussian
distribution, generated by
\texttt{randn(2000, 1) * 0.01} and \texttt{randn(2000, 1) * 0.05}
in \MATLAB.
Even with such non-trivial variations in the central frequencies of
oscillators, the performances do not seem to be affected.

\subsection{A Boolean Logic Example: Half Adder}\seclabel{adder}

As mentioned in \secref{intro}, we can design the Ising Hamiltonian such that
its ground states encode the solutions of a logic circuit.
Then Ising machines can be used to perform invertible Boolean logic computation
\cite{camsari2017pbits,gu2012encoding}.
In this section, we illustrate this capability with a simple example.
We design a small Ising machine that encodes the logic of an adder.
\begin{equation}\eqnlabel{adder}
	a + b = 2c + s,
\end{equation}
where $s$ is the sum and $c$ is the carry bit; all variables are binary, \ie,
$a, b, c, s \in \{0,~1\}$.

The Ising Hamiltonian can be formulated as follows.
\begin{align}
	H &= (a + b - 2c - s)^2 \nonumber\\
      &= a^2 + b^2 + 4c^2 + s^2 + 2ab - 4ac - 2as - 4bc - 2bs + 4cs  \nonumber\\
      &= a + b + 4c + s + 2ab - 4ac - 2as - 4bc - 2bs + 4cs,
\end{align}
where we have used the equality $a^2 = a$ for binary variables.
Such a Hamiltonian function is by definition greater than or equal to zero, and
only reaches zero when the relationship in \eqnref{adder} is satisfied.

We can rewrite the Ising Hamiltonian in a similar format as \eqnref{IsingH0}.
\begin{equation}
	H = \vec h_x^T \cdot \vec x + x^T \cdot \mathbf{J} \cdot x,
\end{equation}
where
\begin{align}
    \vec x &= [c,~s,~a,~b]^T, \\
    \vec h_x &= [4,~1,~1,~1]^T, \\
     \mathbf{J} &= \left[\begin{array}{cccc}
         0 &  2 & -2 & -2 \\
         2 &  0 & -1 & -1 \\
        -2 & -1 &  0 &  1 \\
        -2 & -1 &  1 &  0
         \end{array}\right]. \eqnlabel{adderJ}
\end{align}

To match the definition in \eqnref{IsingH0}, we can convert binary variables
$a, b, c, s$ from $\{0,~1\}$ to $\{-1,~+1\}$ by defining
$\vec s = 2\vec x - 1$. Then we have
\begin{equation}
	H = \vec h^T \cdot \vec s + s^T \cdot \mathbf{J} \cdot s + Const,
\end{equation}
where
\begin{equation}\eqnlabel{hx2h}
    \vec h = \vec h_x - \text{sum}(\mathbf{J},~2) = [2,~1,~-1,~-1]^T,
\end{equation}
with the same $\mathbf{J}$ as in \eqnref{adderJ}.

When this Ising machine settles to the ground state, variables $a, b, c, s$
will satisfy the adder relationship in \eqnref{adder}.
Therefore, if we fix the value of $a$ and $b$, the $c$ and $s$ values we read
from the results will be the carry and sum from the adder circuit.
On the other hand, if $c$ or $s$ is fixed, $a$ and $b$ will settle to values
that solve for the adder relationship.
These predictions can be verified by transient simulation results shown in
\figref{half_adder2} and \figref{half_adder}, generated by the script in
Listing \ref{lst:run_half_adder}.

\begin{figure}[htbp]
	\begin{minipage}{0.49\linewidth}
      \epsfig{file=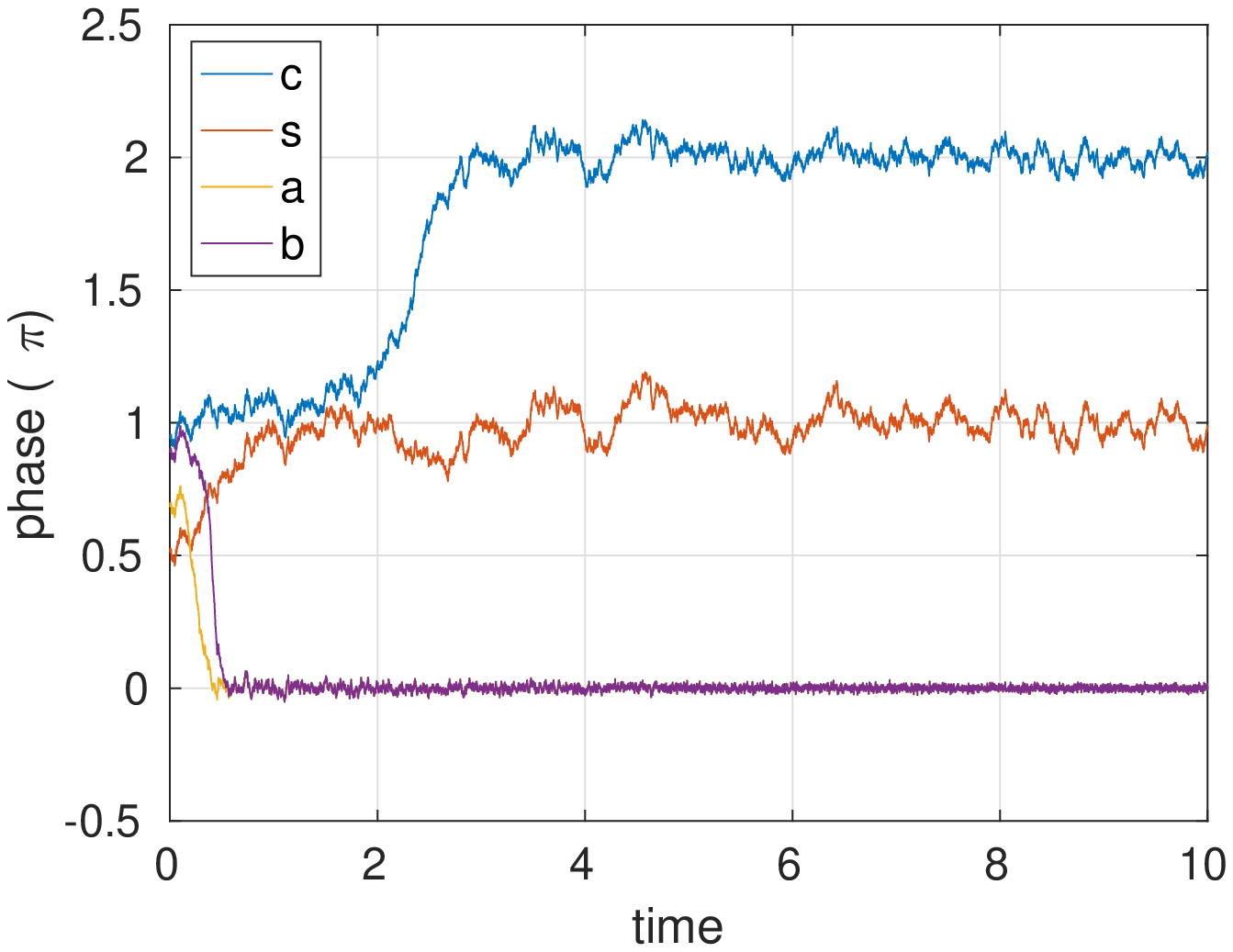,width=\linewidth}
	  \caption{Phase-based half adder ($a + b = 2c + s$) with $a=1$, $b=1$.
               Result here shows $c=1 (2\pi)$, $s=0 (\pi)$.}
               \figlabel{half_adder2}
	\end{minipage}
    \hfill
	\begin{minipage}{0.49\linewidth}
      \epsfig{file=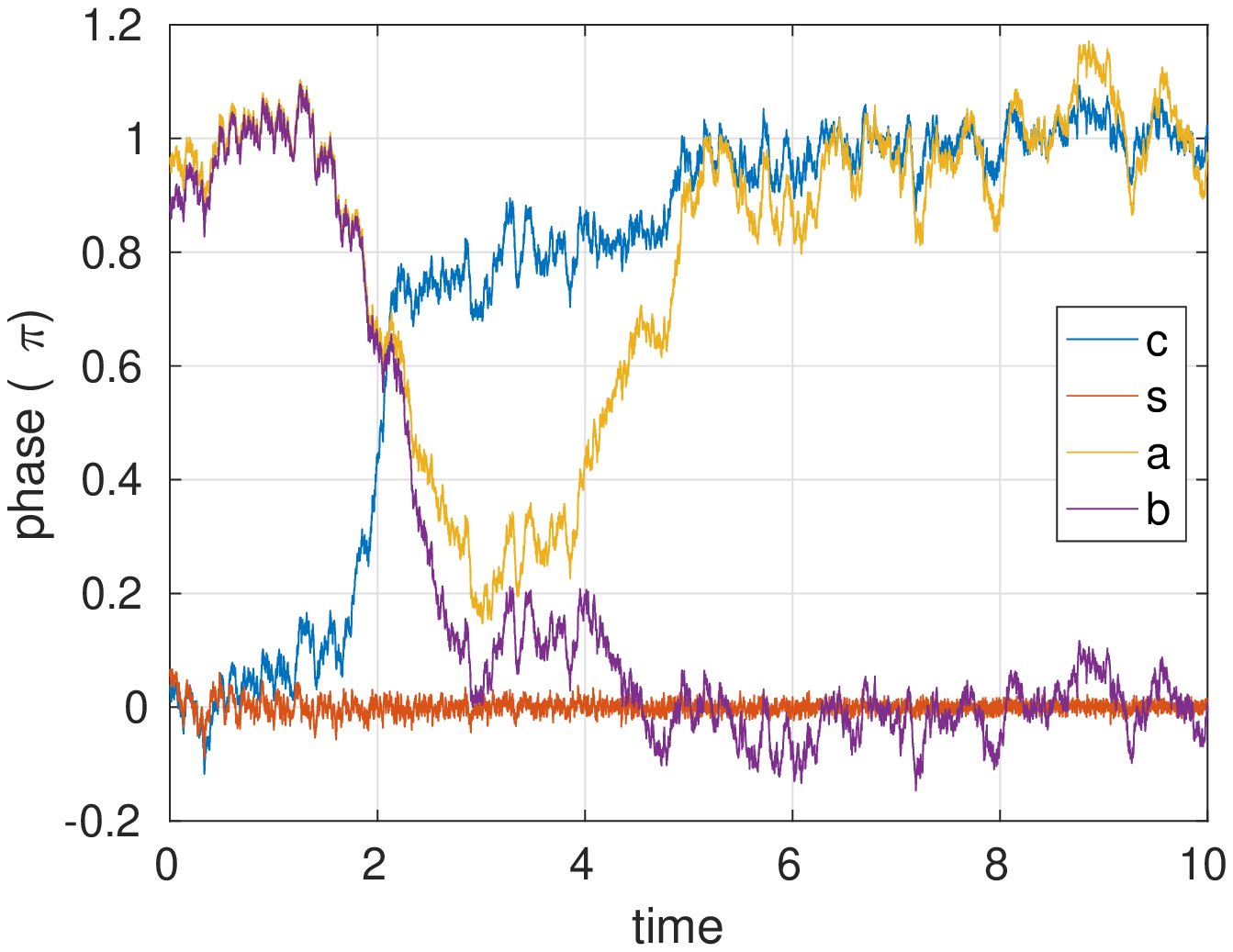,width=\linewidth}
	  \caption{Phase-based half adder ($a + b = 2c + s$) with $s=1$.
               Result here shows $c=0$, $a=0$, $b=1$.}
               \figlabel{half_adder}
	\end{minipage}
\end{figure}

\matlabscript{code/run_half_adder.m}{lst:run_half_adder}{\texttt{run\_half\_adder.m}}

Note that for larger problems, Ising machines normally settle to local optima.
While local optima are often good enough for combinatorial optimization, they
can be meaningless for invertible Boolean logic problems.
Therefore, the suitability of applying Ising machines to logic problems still
needs more study.

\section*{\normalfont {\large Conclusion}}

In this paper, we proposed schemes for implementing Ising machines using
self-sustaining nonlinear oscillators.
We have conducted a comprehensive study of their mechanism, from oscillator
DAEs to the phase macromodel, then to the ``energy'' represented by the
Lyapunov function and its relationship with the Ising Hamiltonian, finally to
the use of Boltzmann law and annealing profiles to achieve better minima.
We have also studied the effect of variations, where our scheme has potential
advantages over existing ones thanks to its use of self-sustaining oscillators
and phase-based logic encoding.
We also showed that the computation time for oscillator-based Ising machines
stays mostly constant as problem size grows, and is dependent mainly on the
oscillator technology.
Finally, the validity and feasibility of the scheme are examined by multiple
levels of simulation and proof of concept hardware implementation.
Simulations run on benchmark combinatorial optimization problems show promising
results, matching the state of art in the development of practical Ising
machines.


\renewcommand{\baselinestretch}{0.8}
\let\em=\it
\bibliographystyle{unsrt}
\bibliography{stringdefs,jr,von-Neumann-jr,PHLOGON-jr,tianshi}

\end{document}